\documentclass[traditabstract,letter]{aa}  
\usepackage{graphicx}
\usepackage{txfonts}
\usepackage{rotating,tabularx}
\usepackage{natbib,twoopt}
\usepackage[breaklinks=true]{hyperref} 
\bibpunct{(}{)}{;}{a}{}{,} 
\newcommandtwoopt{\citeads}[3][][]{\href{http://adsabs.harvard.edu/abs/#3}%
  {\citealp[#1][#2]{#3}}}
\newcommandtwoopt{\citepads}[3][][]{\href{http://adsabs.harvard.edu/abs/#3}%
  {\citep[#1][#2]{#3}}}
\newcommandtwoopt{\citetads}[3][][]{\href{http://adsabs.harvard.edu/abs/#3}%
  {\citet[#1][#2]{#3}}} 
\newcommandtwoopt{\citeyearads}[3][][]%
                 {\href{http://adsabs.harvard.edu/abs/#3}{\citeyear[#1][#2]{#3}}} 

\def\kms    {\ifmmode{{\rm \ts km\ts s}^{-1}}\else{\ts km\ts s$^{-1}$}\fi}
\def\lsol   {\ifmmode{{\rm L}_{\odot}}\else{L$_{\odot}$}\fi}
\def\msol   {\ifmmode{{\rm M}_{\odot}}\else{M$_{\odot}$}\fi}
\def\hi     {\ifmmode{{\rm H}{\rm \small I}}\else{H\ts {\scriptsize I}}\fi}
\def\hh   {\ifmmode{{\rm H}_2}\else{H$_2$}\fi}

\def\zsol   {\ifmmode{{\rm Z}_{\odot}}\else{Z$_{\odot}$}\fi}
\def\tex {\ifmmode{{T}_{\rm ex}}\else{$T_{\rm ex}$}\fi}
\def\tmb {\ifmmode{{T}_{\rm mb}}\else{$T_{\rm mb}$}\fi}

\begin{document} 
\title{Exhaustion of the gas next to M31's supermassive black hole}
\author{Anne-Laure Melchior\inst{1}, Fran\c coise Combes\inst{1,2}}
\institute{LERMA, Observatoire de Paris, PSL Research University, CNRS, Sorbonne Universit{\'e}s, UPMC Univ. Paris 06, F-75014, Paris, France\\
  \email{A.L.Melchior@obspm.fr,Francoise.Combes@obspm.fr}
  \and
 Coll\`ege de France, 11, Place Marcelin Berthelot, F-75\,005 Paris, France}\date{Received April 3, 2015; accepted }
\abstract {New observations performed at IRAM Plateau de Bure reveal the absence of molecular gas next to Andromeda's black hole. We derived a $3\sigma$ upper limit on the molecular gas mass of 4300\,M$_\odot$ for the linewidth of 1000\,km\,s$^{-1}$. This is compatible with infra-red observations which reveal a hole in dust emission next to the black hole. Some gas from stellar feedback is expected from the old eccentric stellar disc population, but it is not accreted close to the black hole. This absence of gas explains the absence of stellar formation observed in this region contrary to what is observed next to Sgr A* in the Milky Way. Either the gas has been swallowed by the black hole, or a feedback mechanism has pushed the gas outside the central 1\,pc. Nevertheless, we detect a small clump of gas with a very small velocity dispersion at 2.4$\arcsec$ from the black hole. It is probable that this clumpy gas is seen in projection, as it does not follow the rotation of the disk surrounding the black hole, its velocity dispersion is ten times smaller than the expected velocity gradient and the tidal shear from the black hole requires a gas density for this clump that is not compatible with our observations.}
\maketitle

\section{Introduction} 
Andromeda is a galaxy, which lies in the green valley \citep{2011ApJ...736...84M,2011A&A...526A.155T,2014ApJ...787...63J}.  According to \citet{2016MNRAS.tmp..899B}, it is typically a low ionisation emission-line region (LIER), as first observed by \citet{1971ApJ...170...25R} and discussed by \citet{1996ASPC..103..241H}. 
\citet{2015A&A...578A..74G} discussed that the torus is disappearing in LIER: there is indeed little gas in the inner part of M31 \citep{2000MNRAS.312L..29M,2011A&A...536A..52M,2013A&A...549A..27M}. 
It is the closest external large galaxy where we can explore the mechanisms that quenched the star formation activity. 
Optical ionised gas has been observed by \citet{2013ApJ...762L..29M} next to the black hole in a field of view\footnote{For compatibility reasons, we assume throughout this letter, a distance to M31 of 780 kpc, i.e. 1 arcsec $\sim$ 3.8 pc, following \citet{2000MNRAS.312L..29M}.} of ${5^{\prime\prime}\times 3.5^{\prime\prime}}$, but this emission is weak.
\citet{1985ApJ...290..136J} estimate the ionised gas mass in the bulge  (${10^{\prime}\times 10^{\prime}}$) of the order of $1500\,M_\odot$. 
 It also hosts a very massive black hole of $1.4\times 10^8 M_\odot$ \citep{2005ApJ...631..280B}, but as studied by \citet{2011ApJ...728L..10L}, it is non-active and only murmurs at a level of $10^{-10}\,L_{Edd}$. 
  It hosts very little star formation  of the order of 0.25-0.3\,M$_\odot$\,yr$^{-1}$, mainly located in the 10-kpc ring of the disc
\citep[e.g.][]{2013ApJ...769...55F,2016MNRAS.456.4128R}. Inside the central region  (${10^{\prime}\times 10^{\prime}}$) , no obvious sign of star formation is detected \citep[e.g.][]{2012ApJS..199...37K,2011AJ....142..139A,2016ApJ...826..136A}, beside a central cluster of A stars formed 200\,Myr ago located next to the black hole  (within $1^{\prime\prime}$) \citep{2012ApJ...745..121L}, designated by P3 by \citet{2005ApJ...631..280B}. \citet{2014A&A...567A..71V} estimate the star formation rate on a pixel basis with panchromatic spectral energy distribution modelling. This infrared-based SFR estimated in the central pixel ($36"\times 36"$) is $4\times 10^{-5}\,M_\odot$\,yr$^{-1}$, while an integration over the central 1-kpc radius region correspond to $1.25\times 10^{-3}\,M_\odot$\,yr$^{-1}$. This negligible SFR is much smaller than the value predicted by \citet{2016MNRAS.456.2537R}, considering supernovae remnants expected within the sphere of influence of quiescent supermassive black holes. For M31, a SFR of $0.13\,M_\odot\,yr^{-1}$ is expected in the sphere of influence ($R_{\mathbf SOI}=14\,$pc$=3.7"$) of its supermassive black hole.  A past AGN activity is also expected and the associated molecular torus, if it survives, should have a radius $R_{MT} = 25$\,pc$=6.7^{\prime\prime}$. In parallel, \citet{2007ApJ...668..236C} expect next to the black hole an accumulation of molecular gas (about $10^4$\,M$_\odot$) originating from stellar feed-back.
In \citet{2013A&A...549A..27M}, we estimate a minimum molecular mass of $4.2\times 10^4\,M_\odot$ within $30\arcsec$ from the centre, while about $10^6\,M_\odot$ of gas is expected from stellar feedback \citep[e.g.][]{1981AJ.....86.1312G}. 

In this letter, we present new observations of molecular gas with IRAM Plateau de Bure interferometer (PdBI). We discuss the implications of the non-detection of gas next to the black hole.
\begin{figure}[h]
  \centering 
 \includegraphics[width=0.5\textwidth]{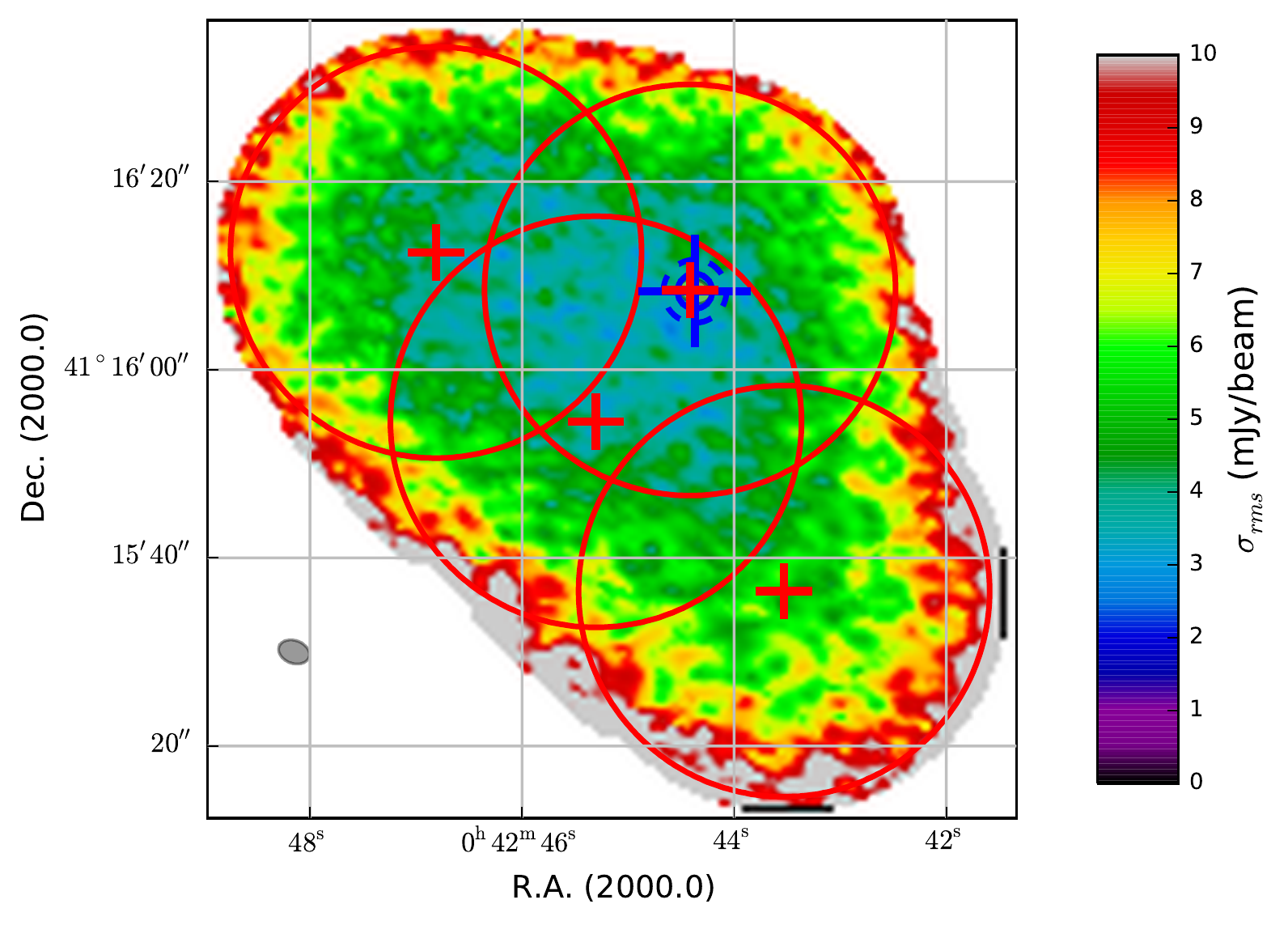}
  \caption{IRAM-PdB CO(1-0) rms map. The colour coding correspond to the noise level ($\sigma_{rms}$), corrected from primary beam, of the map obtained at IRAM-PdB. 
 The blue cross indicates the (optical) centre 
 of Andromeda \citep{1992ApJ...390L...9C}.  
  The red crosses correspond to the centre of the 4 fields observed at PdB with a half-power primary beam of 43.7\arcsec represented by red circles. 
  The half-power beam is presented in the bottom left corner of the CO($1-0$) observations. }
  \label{fig:mympa}%
\end{figure}

\section{Observations}
\begin{figure}[h]
  \centering 
 \includegraphics[width=0.5\textwidth]{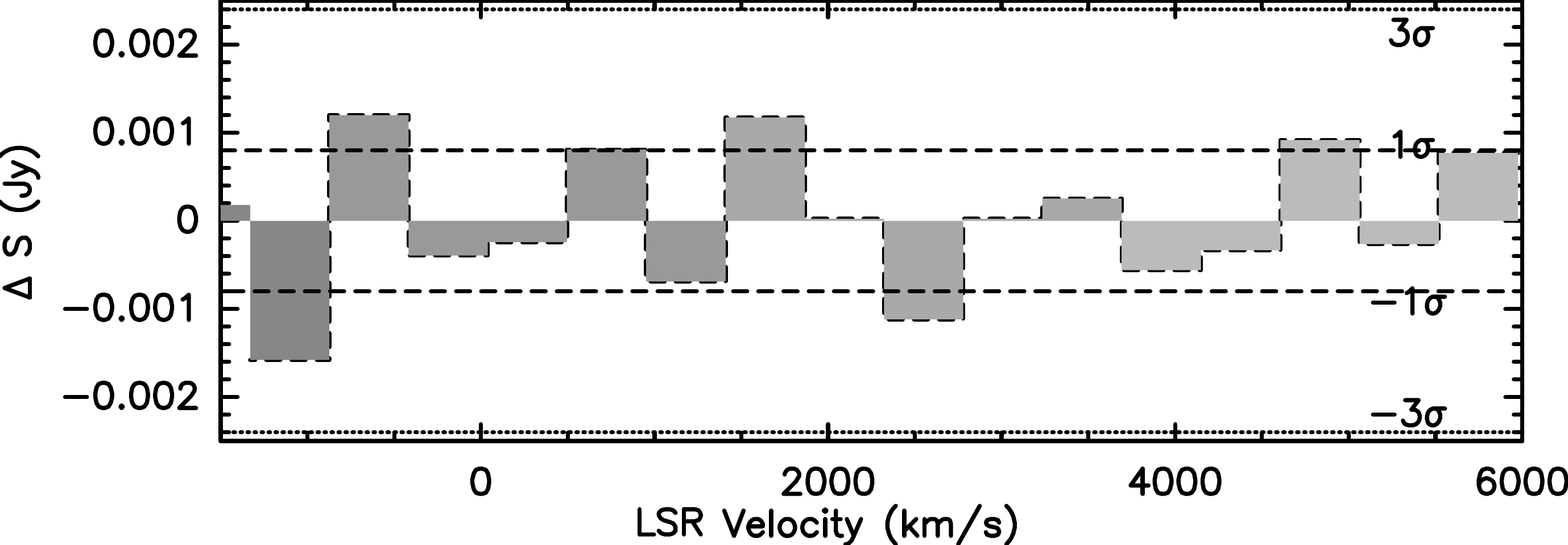}
  \caption{Spectra of the central pixel smoothed at 304\,km/s$^{-1}$. The dashed (resp. dotted) lines correspond to the 1-$\sigma$ (resp. 3-$\sigma$) interval, with $\sigma = 0.6\,$mJy. The systemic velocity of M31 (-300\,km\,s$^{-1}$)  has been removed.}
  \label{fig:centre}%
\end{figure}

The 3-mm observations were carried out at IRAM PdBI with the 5-antenna configuration between 3$^{\mathrm{rd}}$ September and 24$^{\mathrm{th}}$ December 2012 in C-array and D-array. The receivers were tuned to a frequency of 115.386 GHz, to account for M31 systemic velocity (-300\,km/s). We used the WideX correlator providing a broad frequency range of 3.6 GHz and 2 MHz spectral resolution. We thus probe a velocity range of 9000\,km\,s$^{-1}$. The integration time is about 6 hours per field, but for the field at the offset (-16.8$\arcsec$,-21.5$\arcsec$), which has been integrated $34\%$ less: as shown in Figure \ref{fig:mympa}, the level of noise is larger in the South-West part of the data cube. After calibration within the GILDAS reduction package, the visibility set were processed with the MAPPING software. A 1-iteration CLEAN deconvolution was applied, in order to recover a primary-beam corrected data cube close to signal. The beam size of the final data was 3.37"$\times$2.44" (corresponding to 12.8\,pc$\times$9.3\,pc) with a position angle PA$=70\deg$, while the data are sampled with a pixel size of 0.61$\arcsec$. At the black hole position, we reach a sensitivity of $3.2$\,mJy/beam at 1\,$\sigma$ with $\Delta v = 5.1\,km\,s^{-1}$,  and $0.6$\,mJy/beam at 1\,$\sigma$ with $\Delta v = 304\,km\,s^{-1}$, as displayed in Figure \ref{fig:centre}.  The extremities of the band are more noisy and hence, a more optimistic value is found using the radiometer formula with $0.4$\,mJy/beam at 1\,$\sigma$ with $\Delta v = 304\,km\,s^{-1}$.
We subsequently restrict our analysis to the velocity range (-300,+300) km/s. The Figure \ref{fig:mympa} displays the rms map obtained for each channel, where the signal has been removed.

\section{Analysis}
 \begin{table*}
  \caption{Detection of one small clump}             
  \label{table:1}      
  \centering                          
  \begin{tabular}{rrrrrrrrrr}        
    \hline\hline                 
    RA\hspace{1.1cm} DEC  & $\Delta \alpha$ & $\Delta \delta$   &  V$_{LSR}$ & $\Delta V$  & $S_{max}$& $S_{CO} \Delta v$& L$^\prime_{CO}$ (K\,& M$_{H_2}$\\    
    (J2000)\hspace{0.85cm}(J2000)  & ($\arcsec$)& ($\arcsec$)  & (km/s) & (km/s) & (mJy)& (Jy\,km/s)&  km/s\,pc$^2$)& (M$_\odot$) \\    
    \hline                        
    
    00 42 44.52 $+$41 16 10.1 & 1.7  &  1.7  & -176.$\pm$1. & 14.$\pm$3. & 22.2 & 0.32$\pm$0.05 & 480$\pm$70& 2000$\pm$300\\      
     \hline   
                    
                    \\                 
  \end{tabular}
\tablefoot{We provide the J2000 coordinates, offsets, the number of detected pixels, the central velocity V$_{LSR}$, the FWHM $\mathbf{\Delta V}$, the peak intensity $S_{max}$, the integrated line $S_{CO}\Delta v$, the line luminosity $L^\prime_{CO}$ and the molecular mass derived for the clump detected close to the center (in projection). The offsets here and throughout this letter are computed with respect to the optical nucleus (J2000: 00h\,42$^m$\,44.37$^s$ +41$^\circ$\,16$\arcmin$\,8.34$\arcsec$) defined by \citet{1985ApJ...295..287D} (see Figure \ref{fig:phat}).}
\end{table*} 

\subsection{No accumulation of gas next to the black hole}
If there were a gaseous disk/ring surrounding the black hole as observed in the Milky Way \citep[e.g.][]{1998A&A...331..959D,2017A&A...603A..89K} and in external galaxies \citep{2015MNRAS.451.3728R}, given the black hole mass and the stellar velocity dispersion, one would expect a molecular gas signal of 2\,mJy with a linewidth of about 1000\,km\,s$^{-1}$ according to \citet{2007ApJ...668..236C}.
 This correspond to a $10^4$\,M$_\odot$ accumulation of molecular gas next to the black hole due to stellar feed-back from the eccentric inner stellar disc.

Following \citet{2005ARA&A..43..677S}, we can derive the molecular hydrogen mass $M_{H_2}= \alpha_{CO} \times {L^\prime}_{CO}$ with $\alpha_{CO} = 4.36$\, M$\mathbf{_\odot}$\,(K\,km\,s$^{-1}$\,pc$^2$)$^{-1}$, and the line luminosity,  in K\,km\,s$^{-1}$\,pc$^2$:
\begin{equation}
{L^\prime}_{CO} =3.25 \times 10^{7} S_{CO}\Delta V \times \frac{{D_L}^2}{{\nu_{rest}}^2}\frac{1}{(1+z)}
\label{eq:LCO}
\end{equation}
with $S_{CO}\Delta V$ in Jy\,km\,s$^{-1}$, $D_L$ in Mpc and $\nu_{rest}$ in GHz.
Relying on the level of noise we have reached, we estimate an (3\,$\sigma$) upper limit of the molecular gas mass in the beam of 4300\,$M_\odot$ (resp. 2400\,$M_\odot$) for a line width of  1000\,km\,s$^{-1}$ (resp. 300\,km\,s$^{-1}$ ). As discussed above, these values can be increased by 40\% (resp. 30\%) with a direct smoothing of the large band data.

\begin{figure}
    \includegraphics[width=0.5\textwidth]{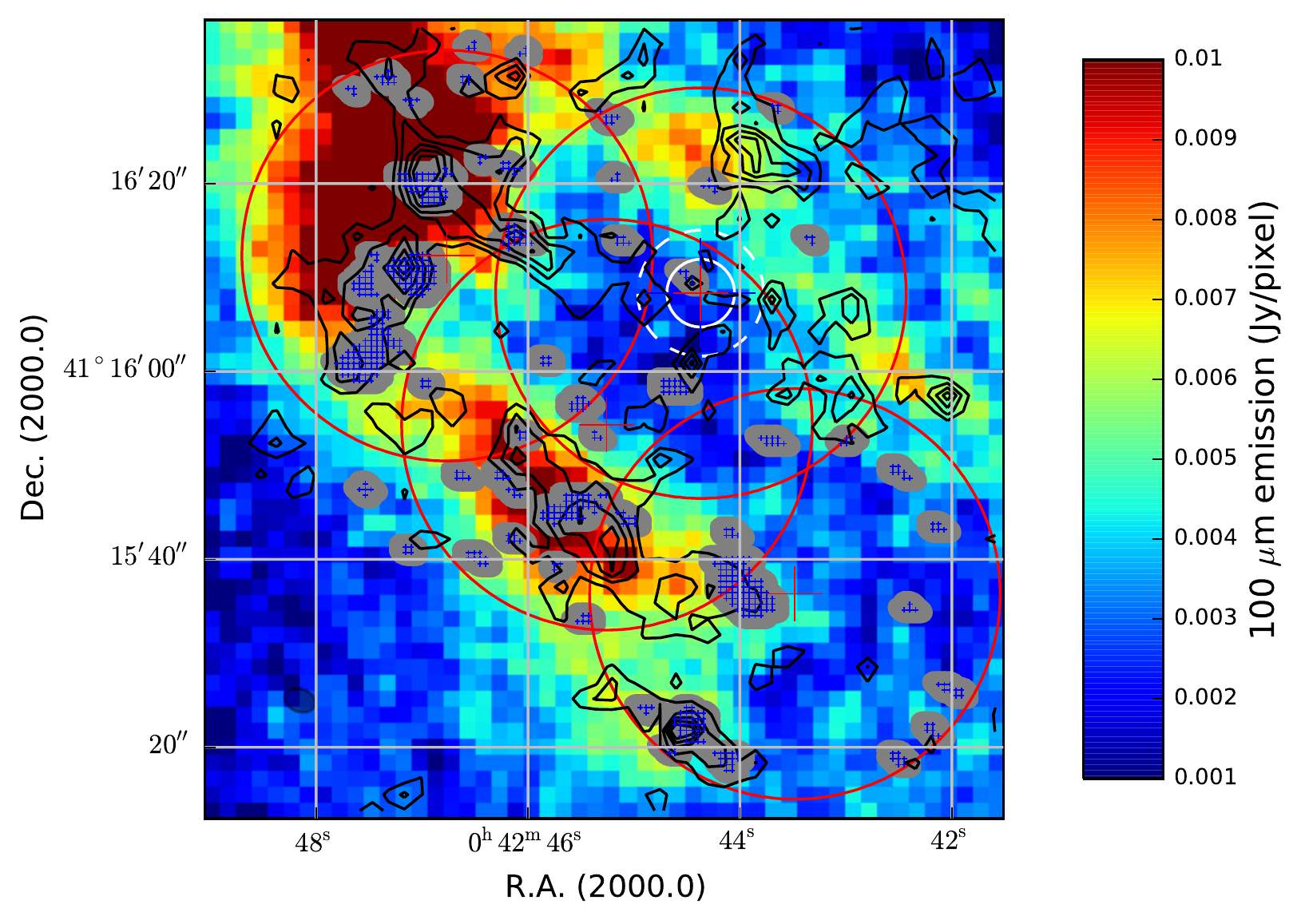}
    \caption{Position of CO excess superimposed on 100$\mu$m Herschel/PACS map. The contours (1.5, 2.6, 3.7, 4.9, 6 MJy/sr) correspond to the Spitzer 8$\mu$m map where the stellar continuum has been substracted by the adjustment of elliptical annuli
using the standard surface photometry algorithm developed for IRAF
\citep{1987MNRAS.226..747J}. The projected field of view is 230\,pc$\times$230\,pc. Following \citet{2016MNRAS.456.2537R}, the dashed-line white circle corresponds to the expected extent of a molecular torus ($R_{MT}=25$\,pc$\sim 6.7\arcsec$), while the full-line white circle corresponds to the radius of the sphere of influence of the black hole ($R_{SOI}=14$\,pc$\sim 3.7\arcsec$), as displayed in Figure \ref{fig:mympa}.  
    } \label{fig:COsources}
\end{figure}

\subsection{Small gas clumps: residuals of a past molecular torus?}
\label{ssect:sgc}
We investigate the possibility that the gas is present under the form of small clumps. Given the good velocity resolution and the quality of the interferometric data, we look for statistically significant weak signals, larger than 3$\sigma$ over a beam area, with the procedure described in Dassa-Terrier, Melchior \& Combes (in prep.). The cores of the detected clumps have been plotted in the Figure \ref{fig:COsources} as blue crosses, while the beam associated with each detected pixel is represented in grey.

We can note that there is an overall correlation with the dust emission detected at 100$\mu$m detected with Herschel/PACS, while CO exhibits a better correlation with the 8$\mu$m dust emission map from Spitzer, displayed as contours in Figure \ref{fig:COsources}.  Next to the centre, the region of interest in this letter, there is no dust emission at 100$\mu$m. Within 12$\arcsec$, there is only 1 clump {with a peak flux above $6\sigma$}, which will be discussed here. The other detections and their reliabilities will be discussed elsewhere (Dassa-Terrier et al., in prep).  
 
\begin{figure}
  \centering 
  \includegraphics[width=0.5\textwidth]{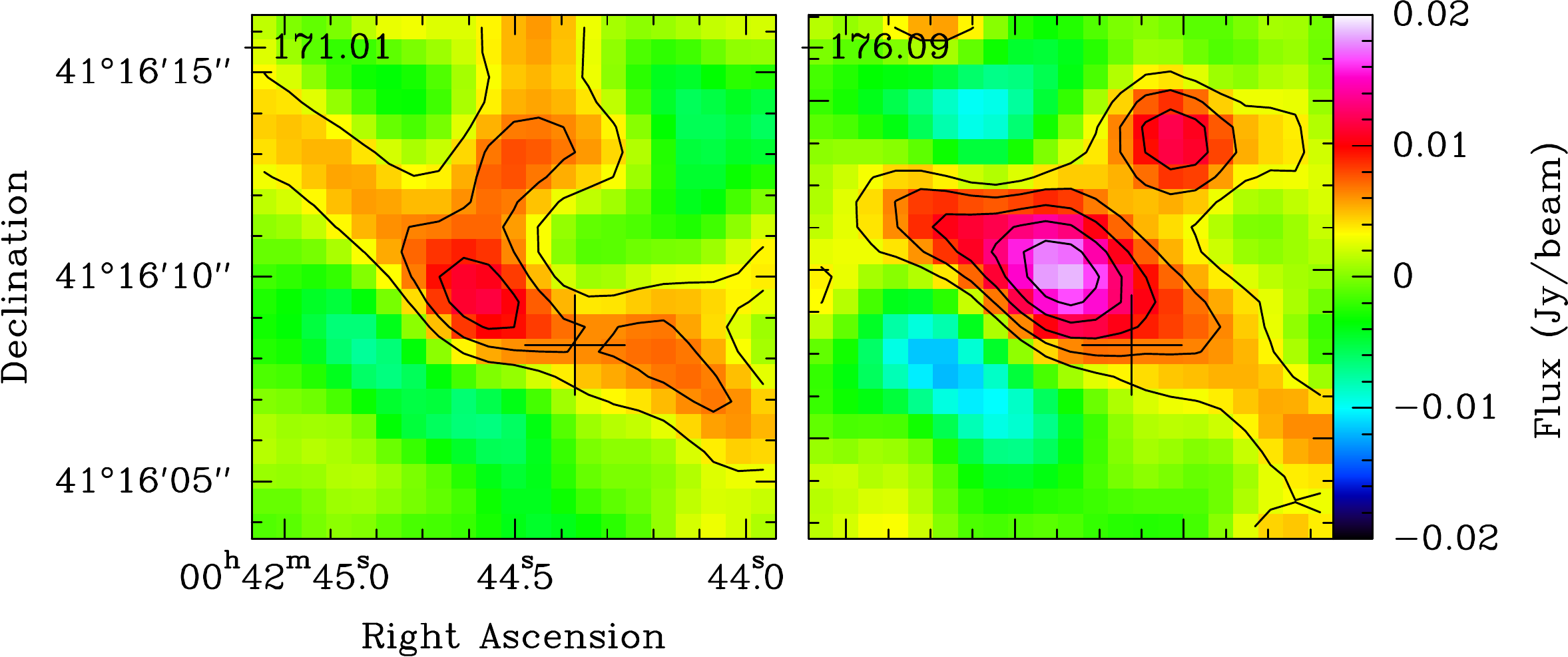}
 \caption{Adjacent channel maps of the clump detected at 2.4$^{\prime\prime}$ (9.1\,pc) from the centre. The contours are represented with a step of 3.2 mJy/beam namely 3.2, 6.4 9.6, 12.8 and 16\,mJy/beam). The cross indicates the position of the optical centre. If it is not an outlier, seen close to the centre only in projection, it lies inside the sphere of influence of the black hole.}
  \label{fig:channels}%
\end{figure}

\begin{figure}
  \centering 
  \includegraphics[width=0.49\textwidth]{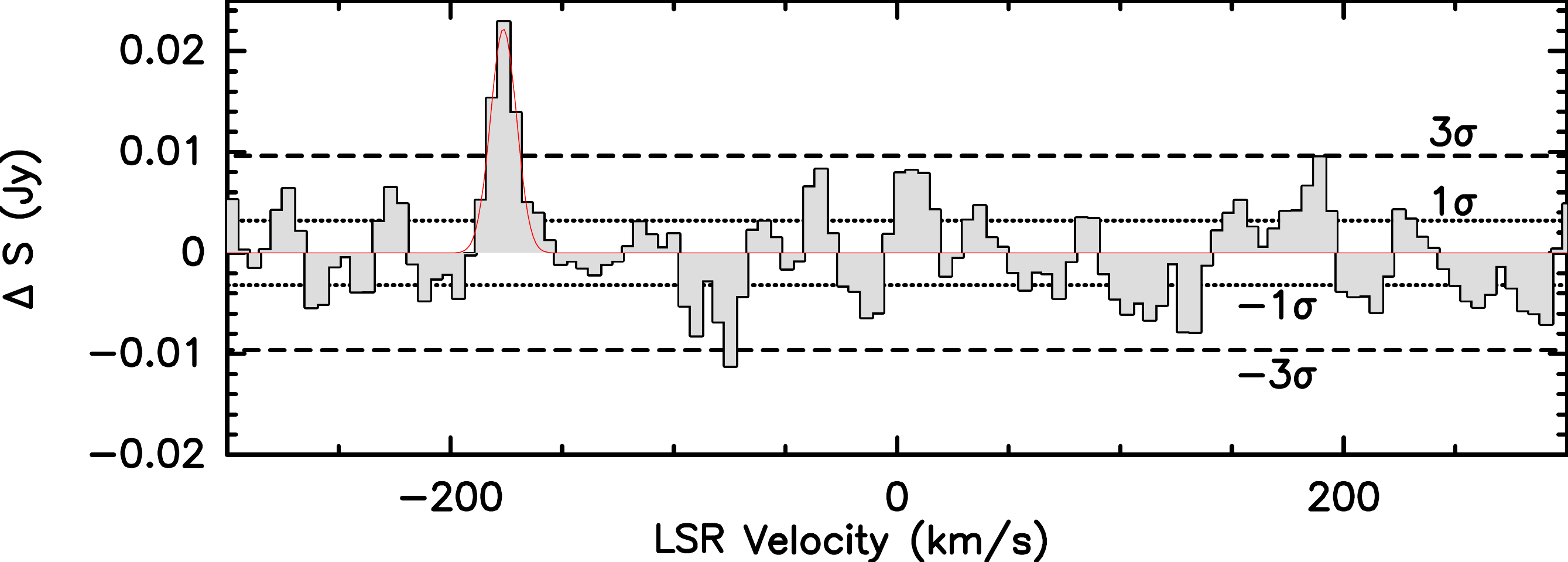}
 \caption{Spectrum of the possible detection at 9.1\,pc,  potentially within the sphere of influence of the black hole.  The systemic velocity of M31 (-300\,km\,s$^{-1}$)  has been removed.}
  \label{fig:spectra}%
\end{figure}

We extract the spectra of the remaining {clump} in {\sc GO VIEW} in MAPPING/Gildas. We then analyse {this} spectra with {\sc CLASS} in Gildas. The results are provided in Table \ref{table:1}.  We estimate a velocity dispersion $\sigma_V \mathbf < 5.9\pm 1.3$\,km/s,  which corresponds to an estimated mass of the clump of 2000\,M$_\odot$. We checked that the signal is not resolved spatially as the integrated line in a single pixel or in the region of interest is about the same within the error bar. {We thus assume they have} a typical projected size of $R_{app}\mathbf < $5.2\,pc {(FWHM/2) corresponding to the beam size or a root-mean-squared (RMS) spatial size of $\sigma_r\mathbf <4.6$\,pc. } Figure \ref{fig:channels} displays the channel map corresponding to this signal, while Figure \ref{fig:spectra} displays its spectra.  In the following, we consider these upper values to discuss qualitative properties of this clump. 
 
According to the Galactic velocity dispersion-size relation of \citet{1987ApJ...319..730S}, {this} clump lies above the $\sigma/(1km/s) = \left( R/(1pc) \right)^{0.5}$ relation. {It has an apparent mass surface brightness 21$M_\odot$/pc$^2$ (computed within the FWHM area). It also lies well above the correlation between $\sigma_v/R^{1/2}$ and $\Sigma$ discussed by \citet{2009ApJ...699.1092H}. }
With these characteristics, one can question if a 2000\,$M_\odot$ clump is gravitationally bound at a distance $D=9.1$\,pc from the black hole. Indeed, it should have a density $n_{tide} = \frac{M_{BH}}{4/3\pi D^3 m_p}>4.2 \times 10^6$\,cm$^{-3}$ to resist the tidal force from the supermassive black hole \citep{2016ApJ...819..138C}. This is much larger than the apparent mean density $n_{app} \sim (M_{H_2}/m_P)/[4/3\pi {R_{app}}^3] \sim 150\,$cm$^{-3}$. This would correspond to a volume filling factor {smaller than} ${3.5 \times 10^{-5}}$, which is well below the volume filling factor ($\sim 1\%$) estimated in \citet{2016A&A...585A..44M} in the gas present along the minor axis in the inner bulge. {This suggests} that the gas is clumped at a smaller scale, as observed in the Galaxy \citep[e.g.][]{2001ApSSS.277...29C,2015ARA&A..53..583H}.  This would correspond to a clump size smaller than $R^\prime_{app} = 0.24$\,pc.  In addition, one can note that the CO luminosity and derived mass are significantly smaller than the main clumps observed in he Central Molecular Zone \citep[e.g.][]{2001ApJ...562..348O}.

 Last, one can consider constraints from the kinematics. A cloud in the gravitational field of the black hole would have a circular velocity of 360\,km/s: if one considers an inclination of 55$\deg$ \citep{2001A&A...371..409B}, one would expect a velocity of 200\,km/s. If there are non-circular motions, this is not incompatible with the amplitude of our value -176\,km/s. However, we would expect a velocity gradient within the beam of about 120\,km/s, while we detect a velocity dispersion of 13\,km/s. Figure \ref{fig:phat} shows the location of the clump discussed here superimposed on an HST colour (336\,nm $-$ 275\,nm) image with respect to the position of the co-called P1, P2, P3 components.

These are convincing signs that this clump is probably not exactly in the mid-plane: if it were at 100\,pc along the line of sight, this gas clump would have the CO critical density and a filling factor of 1$\%$. Then, it would not be in the sphere of influence of the black hole, and its real size and velocity dispersion are probably below our spectral and spatial resolutions.

 \citet{2006PASP..118..590R} discuss that diffuse emission can be missed due to interferometric filtering, but here we expect clumpy gas in this region with little filtering. In addition, these authors discussed that stable recovery of cloud properties can be achieved with a minimum signal-to-noise ratio of 10, while we reach 6.6. It is thus difficult to push more this type of analysis.

\subsection{Comparison with properties derived from infrared data}
Figure \ref{fig:COsources} shows that there is a deficit of 100\,$\mu$m emission next to the centre, in the region surrounding the black hole. Given the lack of resolution of infrared data, dust emission, which seems present in some maps in near and mid-infrared \citep{2012ApJ...756...40S,2015A&A...582A..28P}, corresponds to the emission in the central region as displayed in Figure \ref{fig:COsources}. \citet{2014A&A...567A..71V} estimate a dust mass of 475\,M$_\odot$ in a 36\arcsec$\times$36\arcsec\, pixel, which is 157 times larger that the PdB beam. The stellar continuum is also very strong next to the centre: \citet{2014A&A...567A..71V} compute a stellar mass of 3.71$\times 10^{8}$\,M$_\odot$ in the same region. This central stellar component accounts for the central dust heating as discussed by   \citet{2012MNRAS.426..892G}. This stellar mass corresponds to old stars and should produce some SNIa every 100\,yr \citep{2011MNRAS.412.1473L}. \citet{2001AIPC..565..433S} has detected four supernovae remnants. These supernovae  might maintain supersonic turbulence \citep{2004RvMP...76..125M} and have contributed to expulse the gas from the central region.

\section{Discussion}
\begin{figure}
  \centering 
  \includegraphics[width=0.5\textwidth]{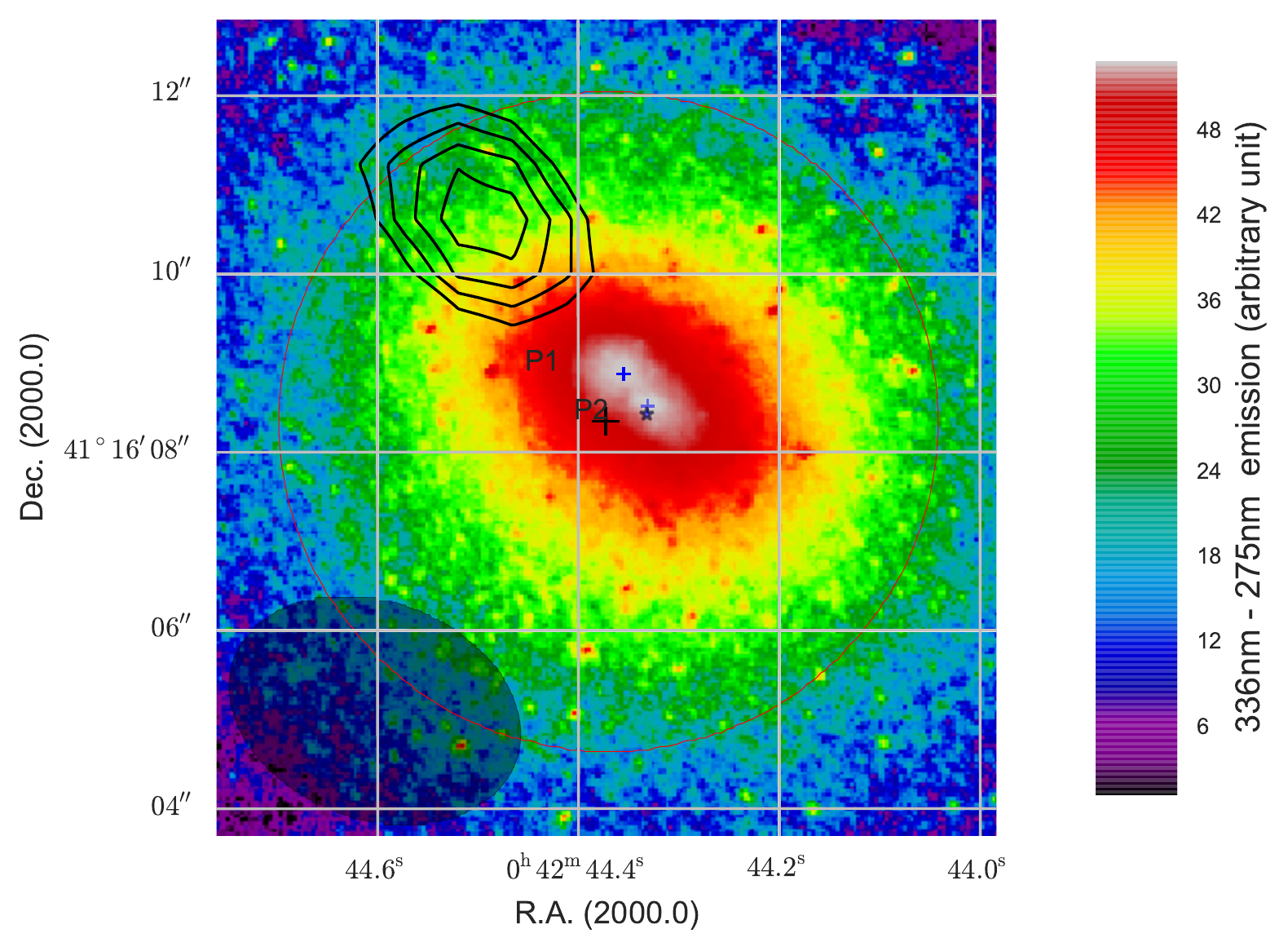}
 \caption{336\,nm  - 275\,nm PHAT map \citep{2012ApJS..200...18D}: the blue crosses correspond to the position of the so-called P1 and P2, the two maxima of this region \citet{2005ApJ...631..280B}. The black cross corresponds to our reference point, quoted by \citet{1992ApJ...390L...9C} as the "optical nucleus" from \citet{1985ApJ...295..287D}. The star symbol refers to the radio nucleus detected by \citet{1992ApJ...390L...9C}: it is in good agreement with the P2 position. On the bottom left, we have the IRAM-PdB beam of our observations. The black contours correspond to our detection within the sphere of influence of the black hole (red circle): we summed the 3 main channels and displayed the contours above 450\,mJy\,km\,s$^{-1}$.}
  \label{fig:phat}%
\end{figure}

We have detected one molecular gas clump with an estimated mass of $2000 M_\odot$, within 2.4$\arcsec$ (9\,pc) from the centre.  Figure \ref{fig:phat} displays a superposition of the central CO clump intensity on the 336\,nm - 275\,nm PHAT/HST image of the central region. While our molecular gas is at the position angle of the P1-P2 disc it is blueshifted, while the stellar disc is redshifted. As discussed in \citet{2011A&A...536A..52M}, it could be rotating in a different orbit, while it could also be not exactly in the mid-plane and belong to a more distant component like the inner disc or inner ring. In addition, as argued in Sect. \ref{ssect:sgc}, the detected velocity dispersion is incompatible with the velocity gradient expected in the gravitation field of a large black hole. We conclude that this clump is outside the sphere of influence of the black hole.

The upper limit we put on the mass of the molecular gas next to the black hole is not compatible with the expectations from stellar mass loss due to the inner eccentric stellar disk. \citet{2007ApJ...668..236C} predicted a gas mass of  ~$10^4\,M_\odot$ due to stellar mass loss and a CO(1-0) signal of 2\,mJy and a linewidth of 1000\,km.s$^{-1}$ within 1\,pc of the black hole. Our observations exclude it at the level of $\mathbf 8.8\sigma$. Given the simultaneous absence of dust emission \citep{2014A&A...567A..71V} and of molecular gas (this work), we can argue that the stellar feed-back produced by the inner eccentric stellar disc population does not accumulate next to the black hole.  Indeed, their model expects a compact source, which should have been detected by our interferometric observations. We can then question whether their proposed mechanism to produce the P3 star cluster is valid. Our result confirms that the current absence of star formation next to the black hole \citep[e.g.][]{2012ApJ...755..131R,2012MNRAS.426..892G} is due to the lack of gas.  As discussed in \citet{2011A&A...536A..52M}, the scenario of a frontal collision with M32 as proposed by \citet{2006Natur.443..832B} could account for the 2-ring morphology as well as for the absence of gas in the centre. In addition, the presence of a bipolar outflow soft X-ray emission, detected along the minor axis by \citet{2007ApJ...668L..39L,2008MNRAS.388...56B} and possibly triggered by supernova  explosions, could account for part of the missing gas.

\begin{acknowledgements}
Based on observations carried out with the IRAM Plateau de Bure Interferometer. IRAM is supported by INSU/CNRS (France), MPG (Germany) and IGN (Spain).
We acknowledge the IRAM Plateau de Bure team for the observations. 
We thank Sabine K{\oe}nig for her support for the data reduction. This project has benefited from supports from \textit{Programme National Cosmologie et Galaxies} and Specific Actions ALMA and \textit{Structuration de l'Univers} from Paris Observatory.  We are most grateful to the anonymous referee for  his/her very constructive comments, which
helped us to substantially improve the manuscript.
A special thanks to Nelson Caldwell for providing a PHAT extract of the central field of Andromeda.
\end{acknowledgements}


\end{document}